\documentclass[letterpaper, 10 pt, conference]{ieeeconf}  

\IEEEoverridecommandlockouts                              
\usepackage{ntheorem}
\usepackage[T1]{fontenc}
\usepackage{mathtools}
\usepackage{fixltx2e}

\makeatletter
\renewcommand*\env@matrix[1][*\c@MaxMatrixCols c]{%
  \hskip -\arraycolsep
  \let\@ifnextchar\new@ifnextchar
  \array{#1}}
\makeatother
\usepackage{graphicx}      
\usepackage{newlfont}
\usepackage{dsfont}
\usepackage{amssymb,amsmath,amsfonts}
\usepackage{epstopdf}
\usepackage{empheq}
\usepackage[dvipsnames]{xcolor}
\usepackage{cuted}
\usepackage{slashbox}

\newcommand{\Real}{{\mathds R}} 
\newcommand{\Nat}{{\mathds N}} 
\newtheorem{definition}{Definition}{}
{}
\newtheorem{proposition}{Proposition}{}
{}
\newtheorem{remark}{Remark}{}
\newtheorem{lemma}{Lemma}{}

\newtheorem{assumption}{Assumption}{}
\newtheorem{problem}{Problem}{}

\makeatletter
\g@addto@macro\normalsize{%
  \setlength\abovedisplayskip{2pt}
  \setlength\belowdisplayskip{2pt}
}
\makeatother

\title{\LARGE \bf
On Privacy of Quantized Sensor Measurements through Additive Noise
}

\author{Carlos Murguia, Iman Shames, Farhad Farokhi, and Dragan Ne\v{s}i\'{c} 
\thanks{This work was partially supported by the Australian Research Council (ARC) under the Discovery Project DP170104099.}
\thanks{Carlos Murguia, Iman Shames, Farhad Farokhi, and Dragan Ne\v{s}i\'{c} are with the Department of Electrical and Electronic Engineering, at the University of Melbourne, Australia.}
\thanks{Emails: carlos.murguia@unimelb.edu.au, \ iman.shames@unimelb.edu.au \ farhad.farokhi@unimelb.edu.au  \& dnesic@unimelb.edu.au.}
}

\begin{document}

\maketitle
\thispagestyle{empty}
\pagestyle{empty}

\begin{abstract}
We study the problem of maximizing privacy of quantized sensor measurements by adding random variables. In particular, we consider the setting where information about the state of a process is obtained using noisy sensor measurements. This information is quantized and sent to a remote station through an unsecured communication network. It is desired to keep the state of the process private; however, because the network is not secure, adversaries might have access to sensor information, which could be used to estimate the process state. To avoid an accurate state estimation, we add random numbers to the quantized sensor measurements and send the sum to the remote station instead. The distribution of these random variables is designed to minimize the mutual information between the sum and the quantized sensor measurements for a desired level of distortion -- how different the sum and the quantized sensor measurements are allowed to be. Simulations are presented to illustrate our results.
\end{abstract}


\section{Introduction}

During the past half-century, scientific and technological advances have greatly improved the performance of engineering systems. However, these new technologies have also led to vulnerabilities within critical infrastructure -- e.g., power, water, transportation. Advances in communication and computing power have given rise to adversaries with enhanced and adaptive capabilities. Depending on adversary's resources and system defenses, opponents may infer critical information about the operation of systems or even deteriorate their functionality. Therefore, designing efficient defence mechanisms is of importance for guaranteeing privacy, safety, and proper operation of critical systems. All these new challenges have attracted the attention of researchers from different fields (e.g., computer science, information theory, control theory) in the broad area of privacy and security of Cyber-Physical Systems (CPS) \cite{Farokhi1}-\nocite{Ahmed2017}\nocite{Farokhi2}\nocite{Murguia2017d}\nocite{RothsteinMorris2017}\nocite{FAROKHI3}\nocite{Murguia2017d}\nocite{Carlos_Justin2}\nocite{Pasqualetti_1}\nocite{Pappas}\nocite{Carlos_Justin1}\nocite{Wyner}\nocite{Ozarow}\nocite{Fawaz}\nocite{Jerome1}\nocite{Hashemil2017}\nocite{Carlos_Justin3}\nocite{Sahand2017}\cite{Carlos_Iman1}. 

In most engineering applications, information about the state of systems is obtained through sensor measurements. Once this information is collected, it is usually quantized, encoded, and sent to a remote station for signal processing and decision-making purposes through communication networks. Examples of such systems are numerous: water and electricity consumption meters, traffic monitoring systems, industrial control systems, and so on. If the communication network is public or unsecured, adversaries might access and estimate the state of the system. To avoid an accurate state estimation, we add random noise to the quantized sensor measurements before transmission and send the sum to the remote station instead. This noise is designed to increase privacy of the transmitted data. Note, however, that it is not desired to overly distort the original sensor data by injecting noise. We might change the data excessively for practical purposes. Hence, when designing the additive noise, we need to take into account the trade-off between \emph{privacy} and \emph{distortion}.

In this manuscript, we follow an information-theoretic approach. We propose to use \emph{mutual information} between quantized-sensor-data and quantized-sensor-data plus privacy noise as \emph{privacy metric}, and the \emph{mean square error} between them as \emph{distortion metric}. The design of the \emph{discrete} additive noise is posed as a convex optimization problem. In particular, the distribution of the noise is designed to minimize the mutual information for a desired level of maximal distortion.

The use of additive noise to increase privacy is common practice. In the context of privacy of databases, a popular approach is differential privacy \cite{Jerome1}-\cite{Dwork}, where noise is added to the response of queries so that private information stored in the database cannot be inferred. In differential privacy, because it provides certain privacy guarantees, Laplace noise is usually used \cite{Dwork2}. However, when maximal privacy with minimal distortion is desired, Laplace noise is generally not the optimal solution. This raises the fundamental question: for a given allowable distortion level, what is the noise distribution achieving maximal privacy? This question has many possible answers depending on the particular privacy and distortion metrics being considered and the system configuration \cite{Topcu}-\nocite{SORIA}\nocite{Geng}\cite{Dullerud}. There are also results addressing this question from an information\linebreak  theoretic perspective, where information metrics -- e.g.,  mutual information, entropy, Kullback-Leibler divergence, and Fisher information -- are used to quantify privacy \cite{Farokhi1}\nocite{Farokhi2}-\cite{FAROKHI3},\cite{Fawaz1}\nocite{Fawaz2}-\cite{Poor}.

In general, if the data to be kept private follows continuous distributions, the problem of finding the optimal additive noise to maximize privacy (even without considering distortion) is hard to solve. If a close-form solution for the distribution is desired, the problem amounts to solving a set of nonlinear partial differential equations which, in general, might not have a solution, and even if they do have a solution, it is hard to find \cite{Farokhi1}. This problem has been addressed by imposing some particular structure on the considered distributions or assuming the data to be kept private is deterministic \cite{Farokhi1},\cite{SORIA},\cite{Geng}.

The authors in \cite{SORIA},\cite{Geng} consider deterministic input data sets and treat optimal distributions as distributions that concentrate probability around zero as much as possible while ensuring differential privacy. Under this framework, they obtain a family of piecewise constant density functions that achieve minimal distortion for a given level of privacy. In \cite{Farokhi1}, the authors consider the problem of preserving the privacy of deterministic databases using constrained additive noise. They use the Fisher information and the Cramer-Rao bound to construct a privacy metric between noise-free data and the one with the additive noise and find the probability density function that minimizes it. Moreover, they prove that, in the unconstrained case, the optimal noise distribution minimizing the Fisher information is Gaussian. 

Most of the aforementioned papers propose optimal continuous distributions assuming deterministic data. However, in a networked context, unavoidable sensor noise leads to stochastic data and thus existing tools do not fit this setting. Here, we identify two possibilities for addressing our problem: 1) we might inject continuous noise to sensor measurements, then quantize the sum, and send it over the unsecured network; or 2), the one considered here, quantize sensor measurements, add noise with discrete distribution, and send the sum over the network. As motivated above, to address the first option, even assuming deterministic sensor data, we have to impose some particular structure on the distributions of the additive noise; and, if sensor data is stochastic, the problem becomes hard to solve (sometimes even untractable). As we prove in this manuscript, if we select the second alternative, under some mild assumptions on the alphabet of the injected noise, we can cast the problem of finding the optimal noise as a constrained convex optimization. To the best of the authors knowledge, this problem has not been considered before as it is posed it here.

\section{Preliminaries}\label{Prelim}

\subsection{Entropy, Joint Entropy, and Conditional Entropy}

Consider a discrete random variable $X$ with alphabet $\mathcal{X}$ and probability mass function $p(x) = \text{Pr}[X=x]$, $x \in \mathcal{X}$, where $\text{Pr}[a]$ denotes probability of event $a$. We denote the probability mass function by $p(x)$ rather than $p_X(x)$ to simplify notation. Thus, $p(x)$ and $p(y)$ refer to two different random variables, and are in fact different probability mass functions, $p_X(x)$ and $p_Y(y)$, respectively.

\vspace{1mm}

\begin{definition}
The entropy of a discrete random variable $X$ with alphabet $\mathcal{X}$ and probability mass function $p(x)$ is defined as $H[X]:= -\sum_{x \in \mathcal{X}}p(x)\log p(x)$.
\end{definition}
The $\log$ is base 2 and thus the entropy is expressed in bits. We use the convention that $0 \log 0=0$ \cite{Cover}.

\begin{definition}
The joint entropy of a pair of discrete random variables $(X,Y)$ with alphabets $\mathcal{X}$ and $\mathcal{Y}$, respectively, and joint probability mass function $p(x,y)$ is defined as $H[X,Y]:= -\sum_{x \in \mathcal{X}}\sum_{y \in \mathcal{Y}}p(x,y)\log p(x,y)$.
\end{definition}

\begin{definition}
Let $(X,Y) \sim p(x,y)$, then the conditional entropy of $Y$ given $X$, $H[Y|X]$, is defined as
\[
H[Y|X]:= -\sum_{x \in \mathcal{X}}\sum_{y \in \mathcal{Y}}p(x,y)\log p(y|x).
\]
\end{definition}

\begin{lemma}\emph{{\cite{Cover}}\hspace{.1mm} \emph{(Chain Rules for Entropy)}}\vspace{1mm}
\[
\begin{array}{ll}
\bullet \hspace{1mm} H[X,Y] = H[X] + H[Y|X].\\[2mm]
\bullet \hspace{1mm} H[X,Y|Z] = H[X|Z] + H[Y|X,Z]. \\[2mm]
\bullet \hspace{1mm} H[Y_1,\ldots,Y_n] = \sum_{i=1}^{n} H[Y_i|Y_{i-1},\ldots,Y_{1}]. \\[2mm]
\bullet \hspace{2mm} $Let $ Z=Z_1,\ldots,Z_m, $ then:$\\[1mm]
\hspace{3mm} H[Y_1,\ldots,Y_n|Z] = \sum_{i=1}^{n} H[Y_i|Y_{i-1},\ldots,Y_{1},Z].
\end{array}
\]
\end{lemma}

\begin{figure*}[t]
  \centering
  \includegraphics[scale=.085]{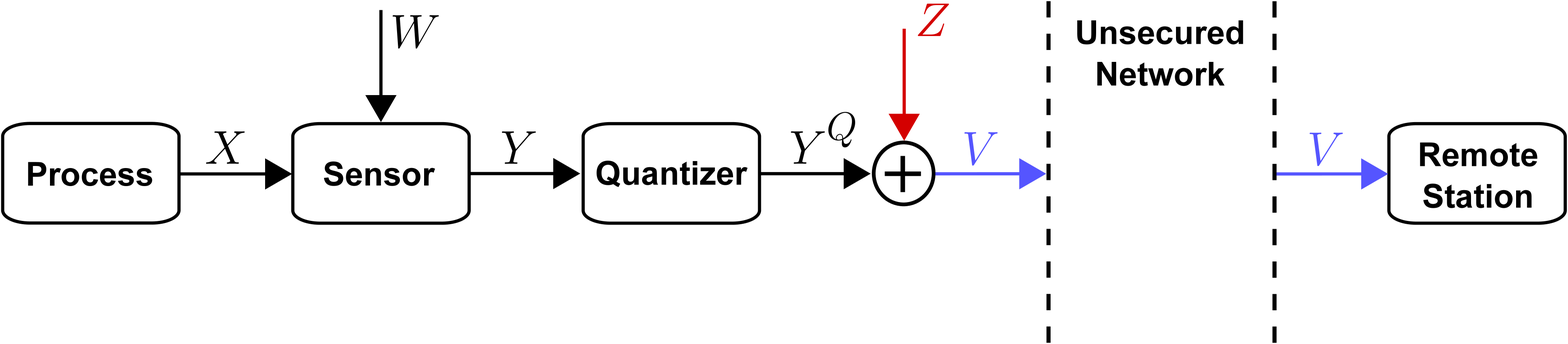}
  \caption{System Configuration.}\label{Fig1}
\end{figure*}

\subsection{Mutual Information}

\begin{definition}
Consider two random variables, $X$ and $Y$, with joint probability mass function $p(x,y)$ and marginal probability mass functions, $p(x)$ and $p(y)$, respectively.\linebreak Their mutual information $I[X;Y]$ is defined as the relative entropy between the joint distribution and the product distribution $p(x)p(y)$, i.e.,
\[
I[X;Y]:= -\sum_{x \in \mathcal{X}}\sum_{y \in \mathcal{Y}}p(x,y)\log \frac{p(x,y)}{p(x)p(y)}.
\]
\end{definition}

\begin{lemma}\emph{{\cite{Cover}} \emph{(Mutual Information and Entropy)}}
\[
\begin{array}{ll}
\bullet \hspace{1mm} I[X;Y] = H[X] - H[X|Y] = H[Y] - H[Y|X].\\[2mm]
\bullet \hspace{1mm} I[X;Y|Z] = H[X|Z] - H[X|Y,Z].\\[2mm]
\bullet \hspace{1mm} $Let $ Z= Z_1,\ldots,Z_m, $ then:$\\[1mm]
\hspace{3mm} I[Y_1,\ldots,Y_n;Z] = H[Y_1,\ldots,Y_n] - H[Y_1,\ldots,Y_n|Z].
\end{array}
\]
\end{lemma}
The mutual information between two jointly distributed random variables, $X$ and $Y$, is a measure of the dependence between $X$ and $Y$. The following properties of mutual information can be found in \cite{Cover} and references therein. Also, sketches of the proofs can be found in \cite{Madiman}.\\[1mm]
\textbf{(P\textsubscript{1})} $I[X;Y] = 0$ if and only if $X$ and $Y$ are independent.\\[1mm]
\textbf{(P\textsubscript{2})} Let $Y$ and $Z$ be independent discrete random variables and $V = Y+Z$; then, $I[V;Y] = H[V] - H[Z]$, i.e., $H[Y+Z|Y]=H[Z]$.\\[1mm]
\textbf{(P\textsubscript{3})} The mutual information does not increase for functions of the random variables (\emph{data processing inequality}):
\[
I[f(X);Y] \leq I[X;Y].
\]

\begin{lemma}\emph{{\cite{Cover}}\hspace{.1mm} \emph{(Chain Rule for Mutual Information)}}\vspace{1mm}
\[
\begin{array}{ll}
\bullet \hspace{1mm} $Let $ Y= Y_1,\ldots,Y_n, $ and $ Z= Z_1,\ldots,Z_m, $ then:$\\[1mm]
\hspace{3mm} I[Y;Z] = \sum_{i=1}^{n} I[Y_i;Y_{i-1},\ldots,Y_{1},Z].
\end{array}
\]
\end{lemma}

\begin{lemma}
Let $Y = Y_1,\ldots,Y_m$ and $Z = Z_1,\ldots,Z_m$ be $2m$ independent discrete random variables and $V = Y+Z$, i.e, $V_i = Y_i+Z_i$, $i=1,\ldots,m$; then:
\[
I[V;Y] = \sum_{i=1}^{m} I[V_i;Y_i] = \sum_{i=1}^{m} H[V_i] - H[Z_i].
\]
\end{lemma}
\emph{\textbf{Proof}}: By Lemma 2, $I[V;Y] = H[V] - H[V|Y]$, and, by Lemma 1, $H[V] - H[V|Y] = \sum_{i=1}^{n} H[V_i|V_{i-1},\ldots,V_{1}]-H[V_i|V_{i-1},\ldots,V_{1},Y]$. By assumption, the elements of $\{Z,Y\}$ are all independent; then, the elements of $V$ are also independent. It follows that
\begin{align*}
H[V] - H[V|Y] &= \sum_{i=1}^{n} H[V_i]-H[V_i|Y]\\
              &= \sum_{i=1}^{n} H[V_i]-H[Y_i+Z_i|Y]\\
              &= \sum_{i=1}^{n} H[V_i]-H[Z_i]\\
              &= \sum_{i=1}^{n} I[V_i;Y_i],
\end{align*}
where the last equality follows from (P\textsubscript{2}) given above. \hfill $\blacksquare$

\section{Problem Setup}
Let $X \in \Real^n$ be the state of some deterministic process that must be kept private. Information about the state is obtained through $m$ sensors of the form:
\begin{equation}\label{sensor_model}
Y = CX + W,
\end{equation}
with sensor measurements $Y \in \Real^m$, matrix $C \in \Real^{m \times n}$, and sensor noise $W \in \Real^m$, $E[W]=\mathbf{0}$, $\Sigma_W := E[W W^T]$, $\Sigma_W > 0$. The entries of the noise are  uncorrelated, i.e., $\Sigma_W =\text{diag}[\sigma_1^2,\ldots,\sigma_m^2]$. Then, $E[Y] = CX$, the covariance $\Sigma_Y := E[(Y-CX)(Y-CX)^T] = \Sigma_W$, and the entries of $Y$ are uncorrelated. We assume that the probability distribution of $Y$ is \emph{known}. This is not an strong assumption since it is often possible to obtain a number of realization of $Y$ to estimate its distribution. Let $Y = (Y_1,\ldots,Y_m)^T$. Each sensor measurement $Y_i$, $i=1,\ldots,m$ is quantized using a uniform quantizer on a finite range  $Q_i(Y_i,y_i^1,\Delta_i,N_i)$:\vspace{2mm}
\begin{equation}\label{quantizer}
Q_i(Y_i,y_i^1,\Delta_i,N_i) :=
\small\left\{
\begin{array}{l}
y^1_i $ if $ Y_i \in (-\infty,y^1_i+\frac{\Delta_i}{2}], \\[2mm]
y^2_i $ if $ Y_i \in (y^1_i+\frac{\Delta_i}{2},y^2_i+\frac{\Delta_i}{2}], \\[2mm]
y^3_i $ if $ Y_i \in (y^2_i+\frac{\Delta_i}{2},y^3_i+\frac{\Delta_i}{2}], \\[2mm]\hspace{25mm} \vdots \\[1.5mm]
y^{N_i}_i $ if $ Y_i \in (y^{N_i-1}_i + \frac{\Delta_i}{2},\infty),
\end{array}
\right.
\end{equation}
where $y^j_i = y^1_i + (j-1)\Delta_i$, $j=1,\ldots,N_i$. Thus, for each sensor, the $N_i$ quantization levels are given by
\[
\mathcal{Y}_i^Q:= \{y^1_i,y^1_i+\Delta_i,\ldots,y^1_i+(N_i-1)\Delta_i \}.
\]
It follows that the vector of quantized sensor measurements $Y^Q := (Y^Q_1,\ldots,Y^Q_m)^T$, $Y^Q_i := Q_i(Y_i,y_i^1,\Delta_i,N_i)$ is determined by the initial quantization level $y^1_i \in \Real$, the quantization step $\Delta_i \in \Real_{>0}$, and the number of intervals $N_i \in \Nat$, $i=1,\ldots,m$. Note that, because we know the distribution of $Y$ and the quantizer, we can always obtain the probability mass function $p(y^Q)$ of $Y^Q$ (and thus also $p(y^Q_i)$ of $Y^Q_i$). Moreover, the alphabet of the \emph{discrete} random variable $Y^Q_i$ is the set of quantization levels $\mathcal{Y}^Q_i$.\vspace{.5mm}

After $Y$ is quantized, a random vector $Z$ is added to $Y^Q$ to obtain $V := Z + Y^Q$. The vector $V$ is transmitted over an unsecured communication network to a remote station, see Fig. 1. Notice that, if we do not add $Z$ to $Y^Q$ before transmission, information about the state is directly accessible through the unsecured network. To minimize this information leakage, we send the sum $V = Z + Y^Q$ to the remote station instead of directly sending $Y^Q$. Note, however, that we do not want to make $Y^Q$ and $Y^Q + Z$ overly different either. By adding $Z$, we might \emph{distort} $Y^Q$ excessively for any practical purposes. Hence, when designing the distribution of $Z$, we need to consider the trade-off between \emph{privacy} and \emph{distortion}. In this manuscript, we propose to use the mutual information between $V = Z + Y^Q$ and $Y^Q$, $I[V;Y^Q]$, as \emph{privacy metric}, and the mean square error, $E[(V-Y^Q)^2]$, as \emph{distortion metric}. Thus, we aim at minimizing $I[V;Y^Q]$ using the probability mass function of $Z$, $p(z)$, as optimization variable subject to $E[(V-Y^Q)^2] = E[Z^2] \leq \epsilon$, for a desired level of distortion $\epsilon \in \Real_{>0}$. In what follows, we formally present the optimization problem we seek to address.

\begin{problem}
For given $Y^Q$ with corresponding $p(y^Q)$ and desired distortion level $\epsilon \in \Real_{\geq 0}$, find the probability mass function $p(z)$ of $Z$ solution of the optimization problem:
\begin{equation} \label{eq:convex_optimization}
\left\{\begin{aligned}
	&\min_{p(z)}\ I[Y^Q+Z;Y^Q],\\
    &\hspace{1mm}\text{\emph{s.t. }} E[Z^2] \leq \epsilon.
\end{aligned}\right.
\end{equation}
\end{problem}

\begin{remark}
Note that if we had access to $Z$ at the other end of the network, and saturation to $Y^Q+Z$ does not occur, we could recover $Y^Q$ exactly from $Z$, and thus cast the optimization problem in \eqref{eq:convex_optimization} without the distortion constraint.
\end{remark}

\begin{remark}
In Problem 1, we could consider individual constraints for the distortion, i.e., $E[Z_i^2] \leq \epsilon_i$, $\epsilon_i \in \Real_{\geq 0}$, $i=1,\ldots,m$, instead of the joint constraint $E[Z^2] \leq \epsilon$. Indeed, if $E[Z_i^2] \leq \epsilon_i$, then $E[Z^2] \leq \sum_{i=1}^m\epsilon_i$.
\end{remark}

\setlength{\tabcolsep}{0.25em} 
{\renewcommand{\arraystretch}{1.5}
\begin{table*}[ht]
\centering
\caption{Probability mass function $p(v_i)$ of $V_i$.}
\label{table2}
\begin{tabular}{|c||c|c|c|c|c|c|c|c|}
\hline
$V_i$  & $v_i^1:=2y_i^1$ & $v_i^2 := 2y_i^1+\Delta_i$ & $\cdots$ & $v_i^{N_i}:=2y_i^1+(N_i-1)\Delta_i$ & $v_i^{N_i+1}:=2y_i^1+N_i\Delta_i$ & $\cdots$ & $v_i^{2N_i-1}:=2y_i^1+2(N_i-1)\Delta_i$ \\ \hline
$p(v_i)$  & $p_{i,1}^V: = p_{i,1}^Y p_{i,1}^Z$ & $ \begin{array}{l} p_{i,2}^V: = p_{i,1}^Y p_{i,2}^Z +\\ p_{i,2}^Y p_{i,1}^Z \end{array}
 $ & $\cdots$ & $ \begin{array}{l} p_{i,N_i}^V := p_{i,N_i}^Y p_{i,1}^Z +\\ p_{i,N_i-1}^Y p_{i,2}^Z +\\ p_{i,N_i-2}^Y p_{i,3}^Z + \cdots +\\ p_{i,1}^Y p_{i,N_i}^Z \end{array}$ & $ \begin{array}{l} p_{i,N_i+1}^V :=  p_{i,N_i}^Y p_{i,2}^Z +\\ p_{i,N_i-1}^Y p_{i,3}^Z + \cdots +\\ p_{i,2}^Y p_{i,N_i}^Z \end{array}$ & $\cdots$ & $p_{i,2N_i-1}^V: = p_{i,N_i}^Y p_{i,N_i}^Z$ \\ \hline
\end{tabular}
\end{table*}
}

\section{Results}

To delimit the solution of Problem 1, we restrict the class of probability mass functions of $Z$. First, we fix the alphabet $\mathcal{Z}_i$ of $Z_i$ -- the $i$-th component of $Z$ -- to be equal to the alphabet $\mathcal{Y}^Q_i$ of $Y^Q_i$, i.e., equal to the quantization levels. This imposes a tractable convex structure on the objective and restrictions, and reduces the optimization variables to the probabilities of each element of the alphabet. The case with arbitrary alphabet leads to a combinatorial optimization pro\-blem where the objective of $\eqref{eq:convex_optimization}$ changes its structure for different combinations. In this manuscript, we do not address this case; it is left as a future work.

Next, note that, because $X$ is deterministic and the covariance matrix $\Sigma_W$ is diagonal, the elements of the vector $Y^Q$ are mutually independent. Then, if we let $Z$ to have independent components, the objective function $I[Y^Q+Z;Y^Q]$ in \eqref{eq:convex_optimization} can be written as follows.

\begin{proposition}
Let the components of $Z$ be mutually independent; then,
$I[Y^Q+Z;Y^Q] = \sum_{i=1}^{m} I[Y_i^Q+Z_i;Y_i^Q]$ and $I[Y_i^Q+Z_i;Y_i^Q] = H[V_i] - H[Z_i]$, $i=1,\ldots,m$.
\end{proposition}
\emph{\textbf{Proof}}: Proposition 1 follows from Lemma 4. \vspace{2mm}

To impose a decoupled structure in the optimization problem, as pointed out in Remark 2, we consider individual constraints for the distortion, i.e., $E[Z_i^2] \leq \epsilon_i$, $\epsilon_i \in \Real_{\geq 0}$, $i=1,\ldots,m$. Then, we can replace \eqref{eq:convex_optimization} by the following $m$ decoupled optimization problems:
\begin{equation} \label{eq:convex_optimizationb}
\left\{\begin{aligned}
	&\min_{p(z_i)}\ H[V_i] - H[Z_i],\\
    &\hspace{1mm}\text{\emph{s.t. }} E[Z_i^2] \leq \epsilon_i, \hspace{2mm}i=1,\ldots,m,
\end{aligned}\right.
\end{equation}
where $p(z_i)$ denotes the probability mass function of $Z_i$ and $\epsilon_i$ is the desired distortion level associated with the mean square error $E[(V_i - Y_i^Q)^2]$.

In what follows, we focus on the solution of \eqref{eq:convex_optimizationb} assuming independence of $Z$ and restricting the alphabet $\mathcal{Z}_i$ of $Z_i$ to be equal to $\mathcal{Y}^Q_i$.

\begin{assumption}
The entries of $Z$ are mutually independent and the alphabet $\mathcal{Z}_i$ of $Z_i$ is equal to the quantization levels $\mathcal{Y}^Q_i$, i.e., it equals the alphabet of $Y^Q_i$.

\end{assumption}
Next, we write $I[Y_i^Q+Z_i;Y_i^Q] = H[V_i] - H[Z_i]$ in \eqref{eq:convex_optimizationb} in terms of $p(y^Q_i)$ and $p(z_i)$. Denote the probabilities of $Y_i^Q$ and $Z_i$ as follows:
\begin{align}\label{probabilities}
p_{i,j}^Y &:= \text{Pr}[Y_i^Q=y_i^j],\\
p_{i,j}^Z &:= \text{Pr}[Z_i=y_i^j],
\end{align}
with $j=1,\ldots,N_i$. Then, the entropy $H[Z_i]$ is given by $H[Z_i] = -\sum_{j=1}^{N_i} p_{i,j}^Z \log p_{i,j}^Z$ and $E[Z_i^2] = \sum_{j=1}^{N_i} (y_i^j)^2 p_{i,j}^Z$. Moreover, since $y_i^j = y^1_i + (j-1)\Delta_i$, then, in terms of the quantizer parameters,  $E[Z_i^2] = \sum_{j=1}^{N_i} (y^1_i + (j-1)\Delta_i)^2 p_{i,j}^Z$. To get an expression for $H[V_i]$, we need the probability mass function $p(v_i)$ of $V_i$. We compute all the possible elements of the alphabet of $V_i = Y_i^Q+Z_i$ and their corresponding probabilities in terms of the elements of the alphabet $\mathcal{Y}^Q_i$, $y_i^j = y^1_i + (j-1)\Delta_i$. Thus, the random variable $V_i$ has an alphabet with $2N_i-1$ elements and the corresponding probabilities are the sums of the probabilities of equal elements. The probability mass function $p(v_i)$ of $V_i$ is given in Table \ref{table2}.

Now, we can write an explicit expression for the objective function in \eqref{eq:convex_optimizationb}:
\begin{align}\label{objective}
I[Y_i^Q+Z_i;Y_i^Q] &= H[V_i] - H[Z_i],\\
                   &= -\sum_{j=1}^{2N_i-1} p_{i,j}^V \log p_{i,j}^V + \sum_{j=1}^{N_i} p_{i,j}^Z \log p_{i,j}^Z,  \notag
\end{align}
where
\begin{align}\label{E8}
v_i^j &:= 2y_i^1 + (j-1)\Delta_i, \hspace{1mm} j=1,\ldots,2N_i-1,\\[1mm]
p_{i,j}^V &:= \text{Pr}[V_i=v_i^j] \label{E9} \\[1mm]
&= \small \left\{ \begin{array}{l}
\sum\limits_{k=1}^{j} p_{i,j+1-k}^Y p_{i,k}^Z, \hspace{1mm} j=1,\ldots,N_i, \\[1mm]
\sum\limits_{k=j+1-N_i}^{N_i} p_{i,j+1-k}^Y p_{i,k}^Z, \hspace{1mm} j=N_i+1,\ldots,2N_i-1.
\end{array} \notag \normalsize  \right.
\end{align}\\
The expressions in \eqref{objective}-\eqref{E9} give a complete characterization of the objective $I[Y_i^Q+Z_i;Y_i^Q]$ in terms of the \emph{known} probabilities of the quantized sensors $p_{i,j}^Y$, $j=1,\ldots,N_i$, and the \emph{optimization variables}, the probabilities of the injected noise $p_{i,j}^Z$, $j=1,\ldots,N_i$. Moreover, the distortion constraint $E[Z_i^2] = \sum_{j=1}^{N_i} (y^1_i + (j-1)\Delta_i)^2 p_{i,j}^Z \leq \epsilon_i$ is linear in $p_{i,j}^Z$. Therefore, if the objective is convex, we could, in principle, efficiently solve \eqref{eq:convex_optimizationb} numerically. However, since $I[Y_i^Q+Z_i;Y_i^Q]  = H[V_i] - H[Z_i]$, $H[V_i]$ is concave in $p_{i,j}^V$, and $H[Z_i]$ is concave in $p_{i,j}^Z$ \cite{Cover}, it is not clear whether $H[V_i] - H[Z_i]$ is convex in $p_{i,j}^Z$ or not.

\begin{proposition}
For given $p_{i,j}^Y$, the function $I[Y_i^Q+Z_i;Y_i^Q]$ is convex in the probabilities $p_{i,j}^Z$, $j=1,\ldots,N_i$.
\end{proposition}
\emph{\textbf{Proof}}: Define the sum:
\begin{align*}
f^V_i:&= - \sum_{j=2}^{2N_i-2} p_{i,j}^V \log p_{i,j}^V.
\end{align*}
The entropy of $V_i$ can be written in terms of $f^V_i$ as
\begin{align}
H[V_i]&= f^V_i - p_{i,1}^V \log p_{i,1}^V - p_{i,2N_i-1}^V \log p_{i,2N_i-1}^V,\label{fV} \\[2mm]
      &= f^V_i - p_{i,1}^Y p_{i,1}^Z \log p_{i,1}^Y - p_{i,1}^Y p_{i,1}^Z \log p_{i,1}^Z\notag \\[1.5mm]
      &\hspace{4mm}- p_{i,N_i}^Y p_{i,N_i}^Z \log p_{i,N_i}^Y - p_{i,N_i}^Y p_{i,N_i}^Z \log p_{i,N_i}^Z \notag ,
\end{align}
where the last equality follows from \eqref{E9}. Write the entropies $H[Z_i]$ and $H[Y_i^Q]$ as
\begin{align}
H[Z_i]   &= - \underbrace{\big(p_{i,1}^Y+\ldots+p_{i,N_i}^Y \big)}_{=1} \sum_{j=1}^{N_i} p_{i,j}^Z \log p_{i,j}^Z,\label{fZ}\\
         &=: f^Z_i - p_{i,1}^Y p_{i,1}^Z \log p_{i,1}^Z - p_{i,N_i}^Y p_{i,N_i}^Z \log p_{i,N_i}^Z,\notag\\
H[Y_i^Q] &= - \underbrace{\big(p_{i,1}^Z+\ldots+p_{i,N_i}^Z \big)}_{=1} \sum_{j=1}^{N_i} p_{i,j}^Y \log p_{i,j}^Y,\label{fY} \\
         &=: f^Y_i - p_{i,1}^Y p_{i,1}^Z \log p_{i,1}^Y - p_{i,N_i}^Y p_{i,N_i}^Z \log p_{i,N_i}^Y.\notag
\end{align}
Combining \eqref{fV}-\eqref{fY}, we can write
\[
H[Z_i] + H[Y_i^Q] = f^Z_i + f^Y_i + H[V_i] - f^V_i,
\]
which implies $H[V_i] = H[Z_i] + H[Y_i^Q] + f^V_i - f^Z_i - f^Y_i$ and thus $I[Y_i^Q+Z_i;Y_i^Q]  = H[Y_i^Q] + f^V_i - f^Z_i - f^Y_i$. The entropy $H[Y_i^Q]$ is constant; then, $I[Y_i^Q+Z_i;Y_i^Q]$ is convex if and only if $f_i(p_{i,1}^Z,\ldots,p_{i,N_i}^Z) := f^V_i - f^Z_i - f^Y_i$ is convex. Next, collecting the $p_{i,j}^Z$ terms and using properties of logarithmic functions, we can write $f_i(p_{i,1}^Z,\ldots,p_{i,N_i}^Z)$ as follows
\begin{equation*}
f_i = \left\{
\begin{array}{l}
\sum\limits_{k=1}^{j} p_{i,k}^Y p_{i,j+1-k}^Z \log \left( \dfrac{p_{i,k}^Y p_{i,j+1-k}^Z}{\sum_{l=1}^{j}p_{i,l}^Y p_{i,j+1-l}^Z} \right),\\[6mm] $\hspace{.1mm} for $j=2,\ldots,N_i, \\[4mm]
\sum\limits_{k=j-1-N_i}^{N_i} p_{i,k}^Y p_{i,j+1-k}^Z \log \left( \dfrac{p_{i,k}^Y p_{i,j+1-k}^Z}{\sum_{l=1}^{j}p_{i,l}^Y p_{i,j+1-l}^Z} \right),\\[6mm] $\hspace{.1mm} for $j=N_i+1,\ldots,2N_i-2.
\end{array}
\right.
\end{equation*}
Note that every element of $f_i(p_{i,1}^Z,\ldots,p_{i,N_i}^Z)$ above is a function of the form $g(a,b,c,\ldots,r) = a\log \big(\frac{a}{a+b+c+\cdots+r} \big)$, $a,b,c\ldots,r \in [0,1]$. The function $g(a,b,c,\ldots,r)$ can be proved to be convex using Theorem 2.7.1 in \cite{Cover} -- \emph{the log sum inequality}. Hence, $f_i(p_{i,1}^Z,\ldots,p_{i,N_i}^Z)$ is the sum of convex functions and thus convex as well. \hfill $\blacksquare$

Note that the ultimate goal is to make it hard for adversaries to infer $X$ from $V=Y^Q+Z$. That is, if someone estimates $X$ using the available data at the public network $V$, the estimation $\hat{X}(V)$ should carry less information about $X$ than an estimate $\hat{X}(Y^Q)$ obtained using $Y^Q$ directly. In other words, we want to make $I[\hat{X}(V);\hat{X}(Y^Q)]$ small.

\begin{proposition}
For some functions $h_V,h_Y: \Real^m \rightarrow \Real^m$, let $\hat{X}(V) := h_V(V)$ and $\hat{X}(Y^Q) := h_Y(Y^Q)$ be estimates of $X$ using $V=Y^Q+Z$ and $Y^Q$, respectively. Then, it is satisfied that $I[\hat{X}(Y^Q+Z);\hat{X}(Y^Q)] \leq I[Y^Q+Z;Y^Q]$ for any pair of functions $h_Y(\cdot)$ and $h_V(\cdot)$.
\end{proposition}
\emph{\textbf{Proof}}: The assertion follows from property (P\textsubscript{3}) in Section \ref{Prelim} -- the \emph{data processing inequality} \cite{Cover}.

\begin{remark}
Proposition 3 has a nice interpretation: for any pair of estimators $(\hat{X}(Y^Q+Z),\hat{X}(Y^Q))$ that can be constructed using $Y^Q+Z$ and $Y^Q$, respectively; the mutual information between them is always upper bounded by $I[Y^Q+Z;Y^Q]$ independently of the estimators. This implies that by minimizing $I[Y^Q+Z;Y^Q]$, we are decreasing the information $I[\hat{X}(Y^Q+Z);\hat{X}(Y^Q)]$. Indeed, the tightness of this bound depends on the particular choice of estimators.
\end{remark}

\subsection{Multiple Observations}

In real-time applications, we often have consecutive observations of the variable $X$ in \eqref{sensor_model}, i.e., a system of the form:
\begin{equation}\label{sensor_model2}
Y(t) = CX + W(t), \hspace{1mm} t \in \Nat,
\end{equation}
with different realizations of sensor data $Y(t) \in \Real^m$ and sensor noise $W(t) \in \Real^m$ at each time step $t$. If the noise $W(t)$ is an i.i.d. process (which is the case most of the time) with $\Sigma_W := E[W(t)W(t)^T] =\text{diag}[\sigma_1^2,\ldots,\sigma_m^2]$ and $E[W(t)]=\mathbf{0}$ for all $t$, the time-dependent model \eqref{sensor_model2} can be written as a static one for a finite number of time-steps $M$. That is, we can collect sensor data for a time window of $M$ steps, stack each set of sensor measurements as $\tilde{Y}_M := (Y(1)^T,\ldots,Y(M)^T)^T \in \Real^{Mm}$, and use this stacked vector to produce a stacked system:
\begin{equation}\label{sensor_model3}
\tilde{Y}_M = \tilde{C}_MX + \tilde{W}_M,
\end{equation}
with sensor noise $\tilde{W}_M := (W(1)^T,\ldots,W(M)^T)^T \in \Real^{Mm}$ and stacked matrix $\tilde{C}_M:=(C^T,\ldots,C^T)^T \in \Real^{Mm \times n}$. Because $W(t)$ is an i.i.d. process and $\Sigma_W$ is diagonal, all entries of $\tilde{W}_M$  and $\tilde{Y}_M$ are mutually independent. Hence, we can use the tools described above to design the distribution of a noise vector $\tilde{Z}_M \in \Real^{Mm}$ that minimizes the mutual information $I[\tilde{Y}_M^Q+\tilde{Z}_M;\tilde{Y}_M^Q]$, where $\tilde{Y}_M^Q$ denotes the quantized $\tilde{Y}_M$. Actually, if we let $\tilde{Z}_M := (Z(1)^T,\ldots,Z(M)^T)^T$ and $Z(t)$ be an i.i.d. process with independent entries, it can be proved that $I[\tilde{Y}_M^Q+\tilde{Z}_M;\tilde{Y}_M^Q] = M I[Y^Q+Z;Y]$, where, with abuse of notation, $Y^Q$ and $Z$ denote two random vectors thrown from the distributions of the i.i.d. processes $Y^Q(t)$ and $Z(t)$. That is, the mutual information $I[\tilde{Y}_M^Q+\tilde{Z}_M;\tilde{Y}_M^Q]$ is simply $M$ times $I[Y^Q+Z;Y]$. It follows that, the distribution of $Z(t) = (Z_1(t)^T,\ldots,Z_m(t)^T)^T$ that minimizes $I[\tilde{Y}_M^Q+\tilde{Z}_M;\tilde{Y}_M^Q]$, for arbitrary large $M$, is the solution, $p(z_i)$, $i=1,\ldots,m$, of problem \eqref{eq:convex_optimizationb}, i.e., $Z_i(t) \sim p(z_i)$, $t \in \Nat$ is the optimal solution.

\section{Simulations Results}

Consider system \eqref{sensor_model} with $X = (\pi^2,\pi^2/4)^T$, $C = I_2$, and sensor noise $W = (W_1,W_2)^T$, $W_1 \sim \mathcal{N}(0,\sigma_1^2)$, $\sigma_1^2 = \pi$, $W_2 \sim \mathcal{U}(-a,a)$, $a=\pi^2/40$, $\sigma_2^2 = (1/3)a^2$. Each sensor measurement $Y_i$, $i=1,2$, is quantized using the uniform quantizer \eqref{quantizer} with $y_1^1 = \pi^2 - 3\sigma_1$, $\Delta_1 = 6\sigma_1/N_1$, $N_1 = 11$,\linebreak and $y_2^1 = 9.09a$, $\Delta_2 = 2a/N_2$, $N_2 = 11$. In Figure 2, we show the optimal distribution $p(z_1)$ of $Z_1$ solution of \eqref{eq:convex_optimizationb}, first without the distortion constraint, and then for the distortion levels $\epsilon_1 = 60,40$. The distortion level for the unconstrained case is $E[Z_1^2] = 105.03$. For comparison, we also show the distributions $p(y_1^Q)$ of $Y_1^Q$ and the one of the sum $V_1=Y_1^Q+Z_1$, $p(v_1)$. In Figure 3, we show the corresponding results for sensor 2: the optimal distributions for the unconstrained case, which yields $E[Z_2^2] = 6.10$, and then for the distortion levels $\epsilon_2 = 5.6,5.1$.

\begin{figure}[t]
  \centering
  \includegraphics[scale=.29]{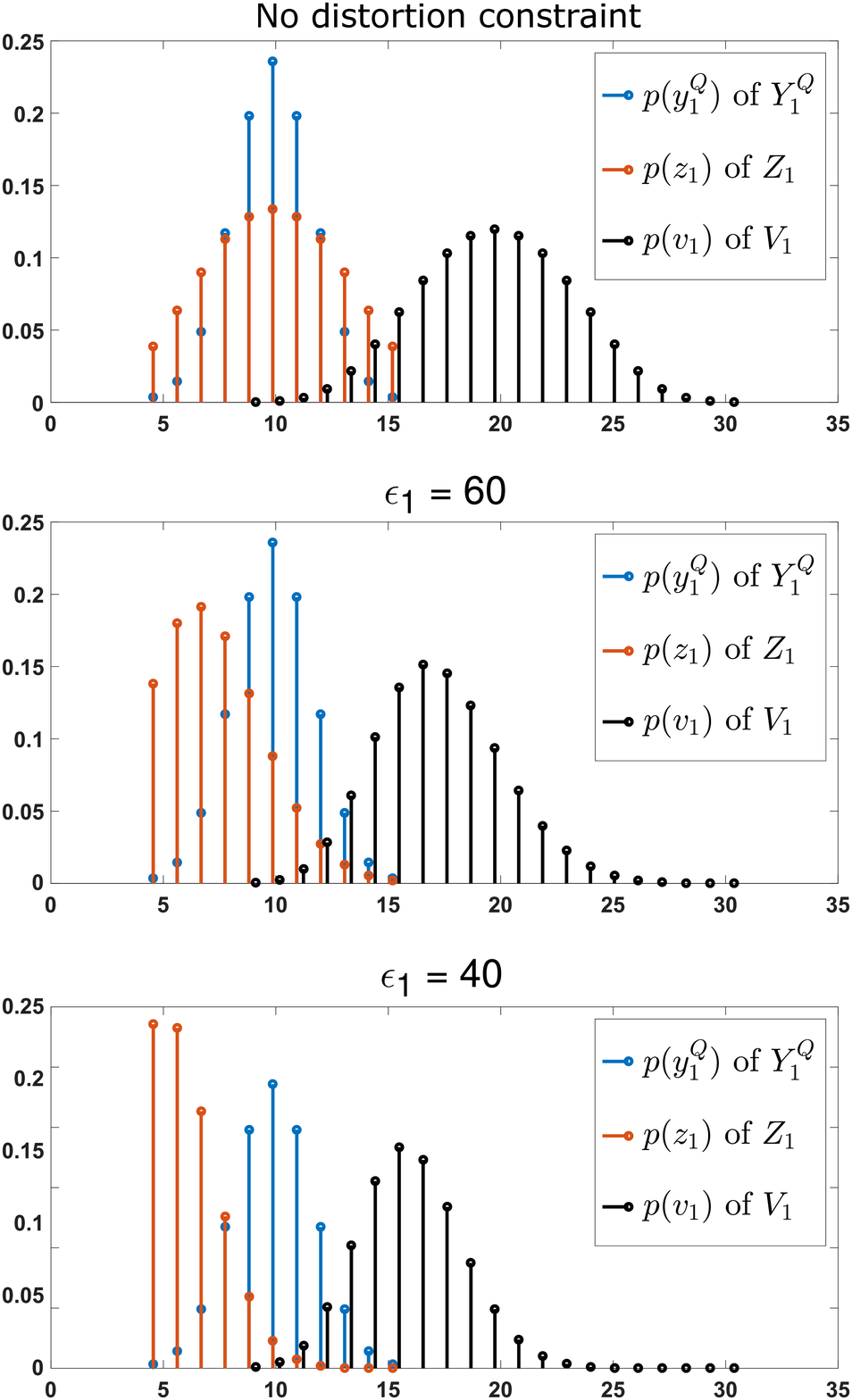}
  \caption{}\label{Fig2}
\end{figure}

\begin{figure}[t]
  \centering
  \includegraphics[scale=.28]{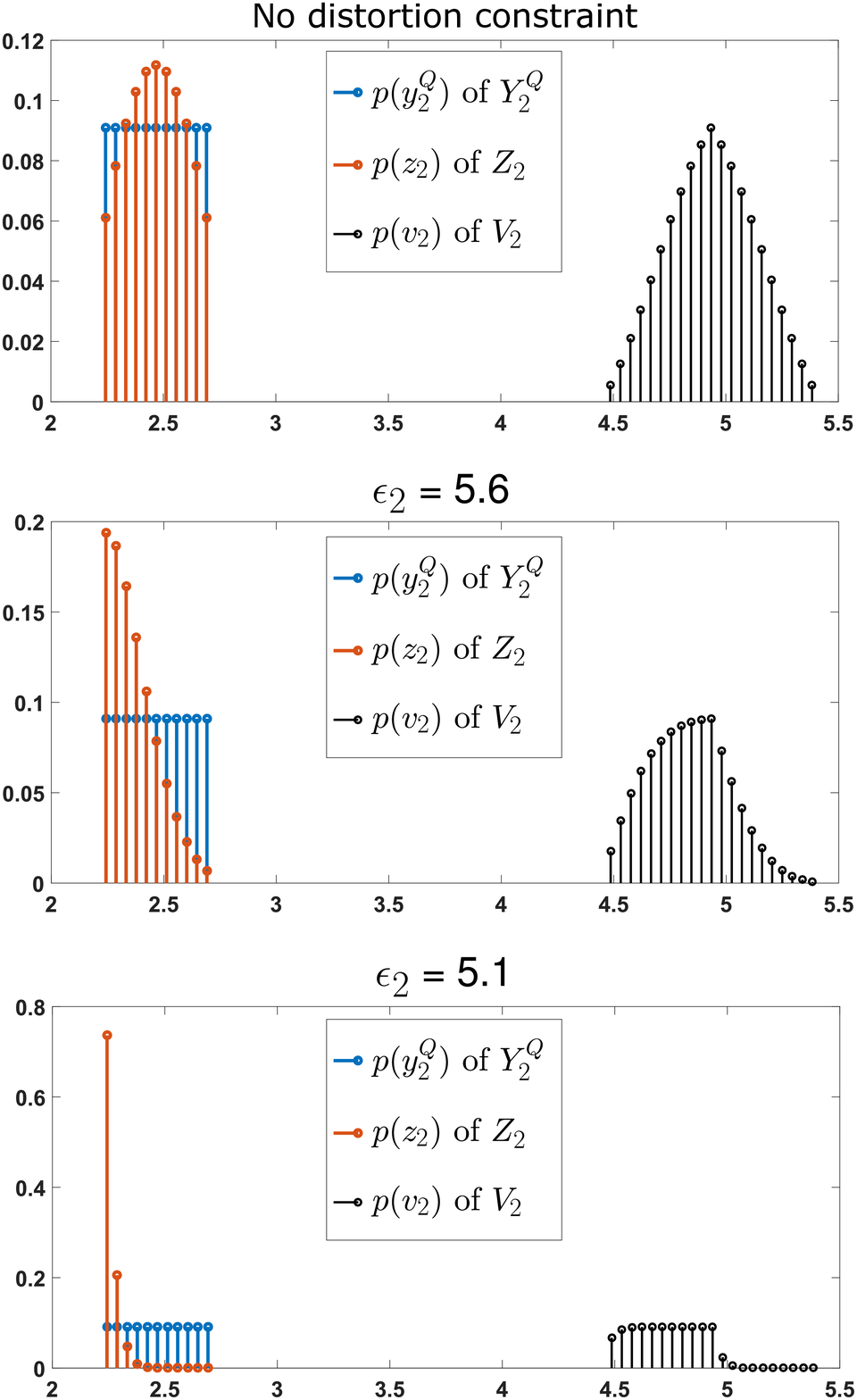}
  \caption{}\label{Fig2}
\end{figure}

\section{Conclusion}

We have provided results on privacy of quantized noisy sensor measurements by adding optimal random variables. To minimize the information leakage due to unsecured communication networks, we have proposed to add random variables to the quantized sensor measurements before transmission. The distributions of these discrete random variables have been designed to minimize the mutual information between the sum and the quantized sensor measurements for a desired level of distortion. In particular, we have posed the design problem as a convex optimization where the optimization variables are the probabilities of the injected noise. We have provided simulation results to test the performance of our tools.





\bibliographystyle{IEEEtran}
\bibliography{ifacconf2}

\end{document}